\documentclass[12pt,a4paper]{article}
\usepackage{amsmath,amssymb,amsfonts}
\usepackage{graphicx}
\usepackage{physics}
\usepackage{bm}
\usepackage{hyperref}

\usepackage{geometry}
\usepackage{float}
\usepackage{booktabs}
\usepackage{siunitx}
\usepackage{tikz}
\usepackage{tikz-3dplot}

\title{Deterministic Reconstruction of Tennis Serve Mechanics: From Aerodynamic Constraints to Internal Torques via Rigid-Body Dynamics}

\author{Sun-Hyun Youn\\[0.5em]
	\textit{Department of Physics, Chonnam National University, Gwangju 61186, Korea}\\[0.3em]
	\texttt{sunyoun@jnu.ac.kr}}

\date{}

\begin{document}
	
	\maketitle
	
	\begin{abstract}
		
		Most conventional studies on tennis serve biomechanics rely on phenomenological observations comparing professional and amateur players or, more recently, on AI-driven statistical analyses of motion data. While effective at describing \textit{what} elite players do, these approaches often fail to explain \textit{why} such motions are physically necessary from a mechanistic perspective. This paper proposes a deterministic, physics-based approach to the tennis serve using a 12-degree-of-freedom multi-segment model of the human upper body. Rather than fitting the model to motion capture data, we solve the inverse kinematics problem via trajectory optimization to rigorously satisfy the aerodynamic boundary conditions required for Flat, Slice, and Kick serves. We subsequently perform an inverse dynamics analysis based on the Principle of Virtual Power to compute the net joint torques. The simulation results reveal that while the kinematic trajectories for different serves may share visual similarities, the underlying kinetic profiles differ drastically. A critical finding is that joints exhibiting minimal angular displacement (kinematically ``quiet'' phases), particularly at the wrist, require substantial and highly time-varying torques to counteract gravitational loading and dynamic coupling effects. By elucidating the dissociation between visible kinematics and internal kinetics, this study provides a first-principles framework for understanding the mechanics of the tennis serve, moving beyond simple imitation of elite techniques.
		
	\end{abstract}
	
	\section{Introduction}
	
	The tennis serve is widely regarded as the most critical stroke in the game, acting as a closed-skill movement where the player has complete control over the initial conditions. Due to its pivotal role, the serve has been a primary subject of sports biomechanics research for decades. Historically, the dominant paradigm in this field has been phenomenological observation. Researchers have extensively utilized high-speed motion capture systems to record the kinematics of elite players, often comparing them with amateur counterparts to identify key differentiators~\cite{elliott2006biomechanics, fleisig2003kinematics}. The prevailing logic of such studies is that the kinematic patterns exhibited by professionals represent the "ideal" form, and deviation from these patterns implies inefficiency or error.
	
	In recent years, with the advent of computer vision and machine learning, this trend has evolved into data-driven approaches. Artificial Intelligence (AI) algorithms can now analyze vast datasets of player movements to identify correlations and predict optimal trajectories~\cite{cust2019machine}. While these methods are powerful in clustering performance metrics and classifying techniques, they often function as "black boxes." They prescribe \textit{what} an optimal motion looks like based on statistical likelihood, but they rarely elucidate \textit{why} such a motion is physically necessary from first principles. For instance, explaining a movement solely because "Top 10 ATP players do it" or "the neural network weighted it highly" lacks the causal mechanical insight required to understand the underlying physics.
	
	The present study diverges from these observational and statistical methodologies by adopting a deterministic, analytical approach based on rigid-body dynamics and Lagrangian mechanics. Instead of mimicking observed phenomena, we aim to derive the serve motion mathematically, treating the human body as a multi-link physical system subject to specific boundary conditions. 
	
	This work builds upon a series of physical analyses of tennis mechanics. Previous studies have established the double pendulum model for stroke dynamics~\cite{youn2015}, analyzed the efficiency of kinetic chains in groundstrokes~\cite{youn2023}, and investigated the optimal swing patterns for target accuracy~\cite{youn2017}. Most recently, the collision mechanics involved in spin generation have been quantified~\cite{youn2025}. 
	
	Extending these foundations, this paper attempts a comprehensive reconstruction of the serve from the preparation phase to the moment of impact. We formulate the problem as a trajectory optimization task: given a desired ball trajectory determined by aerodynamics and court geometry, what are the requisite joint angles and, critically, the joint torques required to achieve this outcome?
	While we acknowledge that real-world implementations vary due to factors such as string tension, ball pressure, and individual physiological constraints~\cite{brody2002physics, cross2011physics, cross2003oblique, goodwill2004ball}, a physics-based model provides a unique advantage. It reveals the invisible kinetics—specifically, the internal torques—that drive the visible kinematics. By solving the inverse dynamics problem, we demonstrate how specific joint torques must evolve to satisfy the physical constraints of the serve, offering a mechanistic explanation that goes beyond mere imitation of professional players.
	
	The remainder of this paper is organized as follows. Section 2 defines the 12-degree-of-freedom kinematic model of the upper body. Section 3 determines the boundary conditions for flat, slice, and kick serves based on realistic aerodynamic trajectories. Section 4 presents the optimization of joint kinematics, and Section 5 performs the inverse dynamics analysis to compute the driving torques, followed by the summary and discussion.

	\section{A Physical Model of the Human Serve: From Core Torque to Distal Velocity}
	
	\subsection{Reference Frames and Generalized Coordinates}
	
	We define an inertial reference frame (World Frame, $\mathcal{F}_W$) fixed to the court, with the origin defined at ground level vertically below the server's initial hip position. The basis vectors are defined as follows: the unit vector $\hat{\mathbf{x}}$ points laterally toward the dominant side (rightward for a right-handed server), $\hat{\mathbf{y}}$ points horizontally toward the net (the primary direction of projection), and $\hat{\mathbf{z}}$ is aligned with the local gravitational vertical.
	
\begin{figure}[htbp]
		\centering
		\includegraphics[width=10cm]{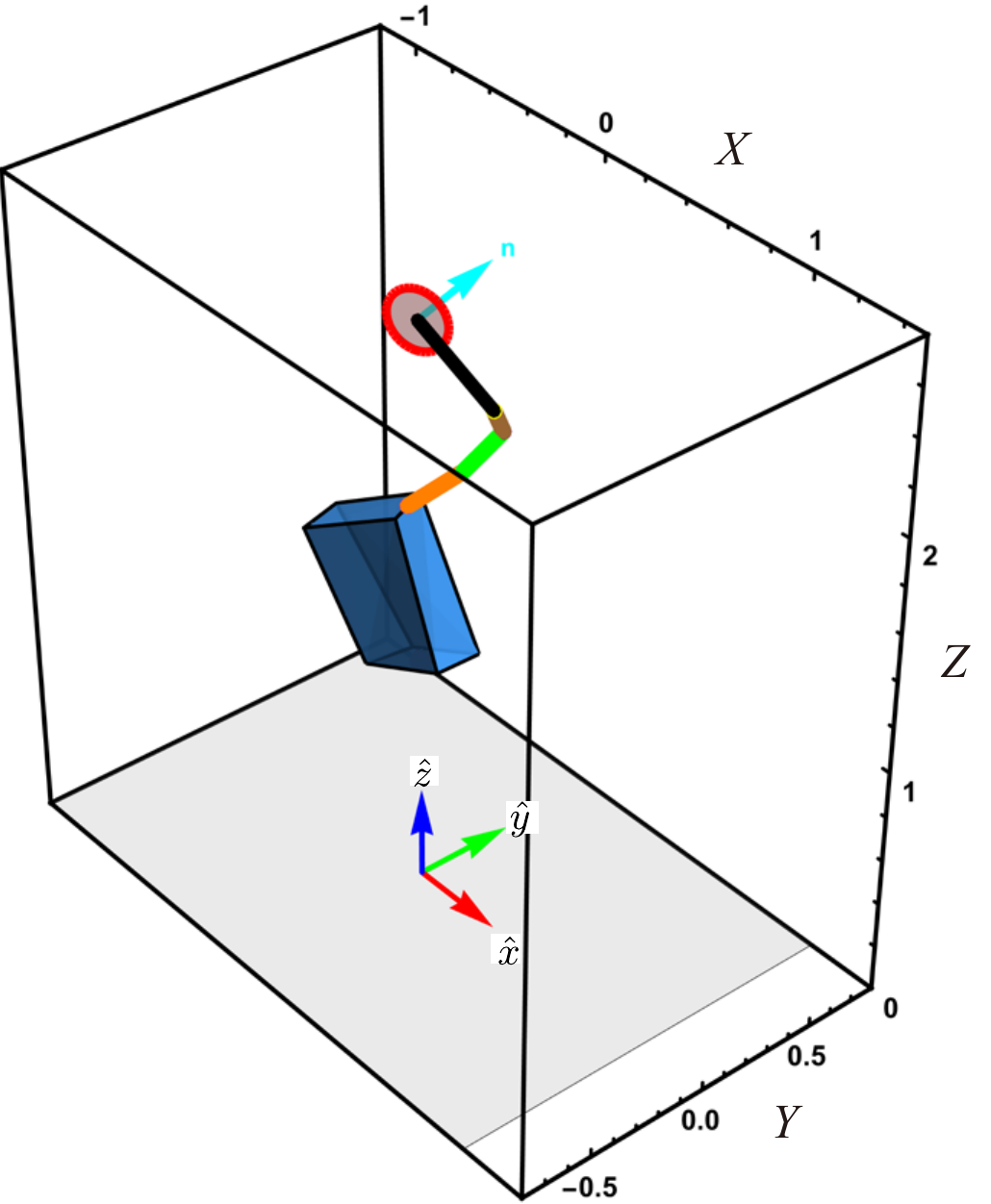}
		\caption{Schematic representation of the multi-segment kinematic model showing the twelve joint angles. The global coordinate system is defined with $x$ pointing laterally toward the dominant side, $y$ pointing toward the net, and $z$ pointing vertically upward. The angular variables illustrated in the figure are 
			$(\theta_1, \theta_2, \ldots, \theta_{12}) = 
			(-25^\circ, 5^\circ, 25^\circ, 20^\circ, -5^\circ, 80^\circ, 
			170^\circ, -20^\circ, -40^\circ, 35^\circ, 50^\circ, 30^\circ)$. } \label{BodyFigure1}
	\end{figure}

	The server is modeled as a system of coupled rigid bodies representing the torso, clavicle, arm, hand, and racket. The configuration of this system is parameterized by a set of generalized coordinates $\mathbf{q} = \{ \theta_1, \theta_2, \dots, \theta_{12} \}$, representing twelve angular degrees of freedom from the trunk to the racket. The hip position $\mathbf{r}_{\text{hip}}$ is treated as an external parameter (it can be adjusted independently to enforce a prescribed impact height). A three-dimensional visualization of the model, defined by these angular parameters, is illustrated in Figure \ref{BodyFigure1}.
	
	\subsection{Three-Dimensional Coordinate Analysis of Human Upper Limb Movements} 
	
	The system is treated as an open kinematic chain rooted at the hip. The spatial configuration is determined by a sequence of coordinate transformations mapping the local body-fixed frame of each component to the global frame $\mathcal{F}_W$.

	The generalized coordinates are assigned as follows:
	\begin{itemize}
		\item \textbf{Torso ($\theta_1, \theta_2, \theta_3$):} Modeled as a deformable rod with lateral flexion ($\theta_1$), anterior-posterior flexion ($\theta_2$), and axial torsion ($\theta_3$). The axial torsion is distributed continuously along the spinal axis.
		\item \textbf{Shoulder Complex ($\theta_4, \theta_5, \theta_6$):} Represents the sternoclavicular and glenohumeral motions, including elevation/depression ($\theta_4$), horizontal abduction/adduction ($\theta_5$), and internal/external rotation about the clavicular axis ($\theta_6$).
		\item \textbf{Distal Arm Segment ($\theta_7, \theta_8$):} Effective flexion/extension of the distal arm segment relative to the shoulder chain ($\theta_7$) and an axial twist of this segment about its longitudinal axis ($\theta_8$). In the present lumped-segment model this pair of DOFs plays the role of elbow flexion and forearm pronation/supination in determining the orientation of the forearm--hand complex.
		\item \textbf{Wrist ($\theta_9, \theta_{10}$):} Flexion/extension ($\theta_9$) and radial/ulnar deviation ($\theta_{10}$) of the hand relative to the distal arm segment.
		\item \textbf{Grip Interface ($\theta_{11}, \theta_{12}$):} Static parameters defining the racket's orientation relative to the hand, where $\theta_{11}$ ($\alpha$) is the grip angle and $\theta_{12}$ ($\beta$) is the face inclination. For a given grip choice these angles are held fixed while the more proximal joint angles are varied.
	\end{itemize}
	
	\subsection{Coordinate Transformations}
	
	The spatial orientation of each rigid body in the chain is described by rotation matrices that transform vectors from the local body-fixed frame to the world frame. We adopt a column-vector convention: a vector expressed in a local frame is premultiplied by the appropriate rotation matrix, and successive rotations are represented by matrix products in which the rightmost factor acts first. Thus the composite transformation from the hand to the world frame has the form
	\[
	\mathbf{R}_{\text{hand}\rightarrow W}
	= \mathbf{R}_{\text{torso}} \, \mathbf{R}_{\text{sh,base}} \, \mathbf{R}_{\text{forearm}} \, \mathbf{R}_{\text{hand}} ,
	\]
	so that $\mathbf{v}_W = \mathbf{R}_{\text{hand}\rightarrow W} \, \mathbf{v}_{\text{hand}}$ for any vector $\mathbf{v}_{\text{hand}}$ expressed in the hand frame.
	
	The elementary rotation operators about the canonical axes are denoted by $\mathbf{R}_x(\phi)$, $\mathbf{R}_y(\phi)$, and $\mathbf{R}_z(\phi)$, and are taken to be active, right-handed rotations.
	
	\subsubsection{Torso and Spinal Deformation}
	The torso is not treated as a single rigid body but rather as a continuum capable of torsion. The base orientation at the hip is determined by flexural rotations:
	\begin{equation}
		\mathbf{R}_{\text{bend}} = \mathbf{R}_x(\theta_2) \, \mathbf{R}_y(\theta_1) ,
	\end{equation}
	where the rightmost factor $\mathbf{R}_y(\theta_1)$ acts first on local torso vectors. To model the spinal kinematics, the axial rotation $\theta_3$ is distributed linearly along the vertical axis $z$. The twist angle $\phi(z)$ at a height $z$ is given by
	\begin{equation}
		\phi(z) = \theta_3 \, \frac{z}{H_{\text{torso}}} .
	\end{equation}
	At the shoulder attachment level ($z = H_{\text{torso}}$) the full axial rotation $\theta_3$ is realized, so the net orientation of the upper torso is
	\begin{equation}
		\mathbf{R}_{\text{torso}} = \mathbf{R}_z(\theta_3) \, \mathbf{R}_{\text{bend}} .
	\end{equation}
	
	\subsubsection{Shoulder Complex and Upper Arm Rotational Degrees of Freedom}
	
	The clavicle orientation relative to the torso is defined by elevation and horizontal abduction. The rotation matrix for the shoulder base $\mathbf{R}_{\text{sh,base}}$ is given by the composition of the torso frame and the local shoulder rotations:
	\begin{equation}
		\mathbf{R}_{\text{sh,base}} = \mathbf{R}_{\text{torso}} \, \mathbf{R}_z(\theta_5) \, \mathbf{R}_y(-\theta_4) .
	\end{equation}
	Note that the negative sign in $\mathbf{R}_y(-\theta_4)$ aligns the mathematical positive rotation with the anatomical definition of elevation.
	
	The kinematics of the humerus involve a finite rotation about the clavicular axis. Defining $\hat{\mathbf{u}}_{\text{clav}}$ as the unit vector aligned with the clavicle in the laboratory frame, the rotation tensor for the upper arm is composed as:
	\begin{equation}
		\mathbf{R}_{\text{upper}} = \mathcal{R}(\hat{\mathbf{u}}_{\text{clav}}, \theta_6) \, \mathbf{R}_{\text{sh,base}} ,
	\end{equation}
	where $\mathcal{R}(\hat{\mathbf{n}}, \theta)$ denotes the rotation operator about an axis $\hat{\mathbf{n}}$ by an angle $\theta$. Explicitly, this operator corresponds to the Rodrigues rotation formula, $\mathcal{R}(\hat{\mathbf{n}}, \theta) = \mathbf{I} + \sin\theta \,\mathbf{K} + (1-\cos\theta)\mathbf{K}^2$, with $\mathbf{K}$ being the generator of rotations about $\hat{\mathbf{n}}$ (i.e., $K_{ij} = -\epsilon_{ijk} n_k$). The vector $\hat{\mathbf{u}}_{\text{clav}}$ is obtained by transforming the shoulder's local basis vector into the world frame.

	\subsubsection{Distal Segmental Chain: Forearm--Wrist--Hand--Racket System}
	
	The distal arm segment is modeled as a single rigid link attached to the upper arm, with two rotational DOFs. The first, $\theta_7$, introduces an effective flexion in the local frame defined by the humerus, and the second, $\theta_8$, introduces an axial twist (corresponding kinematically to forearm pronation/supination) about the longitudinal axis of this segment. The transformation for the distal arm frame is
	\begin{equation}
		\mathbf{R}_{\text{forearm}} = \mathbf{R}_{\text{axial}}(\theta_8, \hat{\mathbf{u}}_{\text{arm}}) \, \mathbf{R}_{\text{upper}} \, \mathbf{R}_{\text{elbow}}(\theta_7) ,
	\end{equation}
	where $\hat{\mathbf{u}}_{\text{arm}}$ is the longitudinal axis of the upper arm after application of $\mathbf{R}_{\text{upper}}$ and $\mathbf{R}_{\text{elbow}}(\theta_7)$ denotes the flexion/extension rotation about the local hinge axis.
	
	The hand orientation is subsequently determined by the wrist rotations:
	\begin{equation}
		\mathbf{R}_{\text{hand}} = \mathbf{R}_{\text{forearm}} \, \mathbf{R}_z(\theta_{10}) \, \mathbf{R}_y(\theta_9) .
	\end{equation}
	
	Finally, the racket's orientation is fixed relative to the hand. Defining the hand's local normal vector $\hat{\mathbf{n}}_{\text{palm}}$ and thumb vector $\hat{\mathbf{d}}_{\text{thumb}}$ in the world frame via $\mathbf{R}_{\text{hand}}$, the racket's heading vector $\hat{\mathbf{v}}_{\text{rkt}}$ and face normal $\hat{\mathbf{n}}_{\text{rkt}}$ are computed by applying the grip rotations $\alpha$ and $\beta$ as axial rotations:
	\begin{align}
		\hat{\mathbf{v}}_{\text{rkt}} &= \mathbf{R}_{\text{axial}}(\alpha, \hat{\mathbf{n}}_{\text{palm}})\, \hat{\mathbf{d}}_{\text{thumb}}, \\
		\hat{\mathbf{n}}_{\text{rkt}} &= \mathbf{R}_{\text{axial}}(\beta, \hat{\mathbf{v}}_{\text{rkt}})\, \hat{\mathbf{n}}_{\text{palm}} .
	\end{align}
	
	\subsection{Position Vectors and Kinematics}
	
	The position of the racket head center, $\mathbf{r}_{\text{head}}$, is determined by the vector sum of the constituent segment vectors. Let $L_{\text{clav}}$, $L_{\text{upper}}$, $L_{\text{hand}}$, and $L_{\text{rkt}}$ denote the lengths of the clavicle, distal arm segment, hand (from wrist to grip), and racket (from handle to head center), respectively.
	
	The trajectory of the racket head in the global frame is given by:
	\begin{equation}
		\mathbf{r}_{\text{head}}(\mathbf{q}) = \mathbf{r}_{\text{hip}} + \mathbf{R}_{\text{torso}}\mathbf{d}_{\text{shoulder}} + L_{\text{clav}}\hat{\mathbf{u}}_{\text{clav}} + L_{\text{upper}}\hat{\mathbf{u}}_{\text{arm}} + L_{\text{hand}}\hat{\mathbf{u}}_{\text{hand}} + L_{\text{rkt}}\hat{\mathbf{v}}_{\text{rkt}} ,
	\end{equation}
	where $\mathbf{d}_{\text{shoulder}}$ is the displacement of the shoulder joint relative to the hip within the torso frame, and $\hat{\mathbf{u}}_{\text{clav}}$, $\hat{\mathbf{u}}_{\text{arm}}$, and $\hat{\mathbf{u}}_{\text{hand}}$ are the unit vectors along the respective segments expressed in the global frame.
	
	The instantaneous velocity of the racket head is the time derivative of the position vector. In the Lagrangian formalism, this can be expressed as the product of the translational Jacobian matrix $\mathbf{J}_v(\mathbf{q})$ and the vector of generalized velocities $\dot{\mathbf{q}}$:
	\begin{equation}
		\mathbf{v}_{\text{head}} = \frac{d}{dt}\mathbf{r}_{\text{head}}(\mathbf{q}) = \sum_{i=1}^{12} \frac{\partial \mathbf{r}_{\text{head}}}{\partial \theta_i} \dot{\theta}_i \equiv \mathbf{J}_v(\mathbf{q}) \cdot \dot{\mathbf{q}} .
	\end{equation}
	
	Similarly, the angular velocity of the racket frame, $\boldsymbol{\omega}_{\text{rkt}}$, which is critical for determining the spin imparted to the ball, is given by the superposition of the angular velocity vectors of all preceding joints~\cite{sprigings1994scientific}:
	\begin{equation}
		\boldsymbol{\omega}_{\text{rkt}} = \sum_{i=1}^{12} \dot{\theta}_i \, \hat{\mathbf{n}}_i ,
	\end{equation}
	where $\hat{\mathbf{n}}_i$ represents the unit vector along the axis of rotation for the $i$-th joint expressed in the global frame.

	\section{Ball Trajectory Analysis}
	
	The trajectory of a tennis ball in flight is determined by gravitational and aerodynamic forces, the latter arising from the interaction between the ball's translational velocity and angular velocity with the surrounding air~\cite{mehta2001sports}. Once the initial velocity vector $\mathbf{v}_b$ and spin vector $\boldsymbol{\omega}$ are specified, the subsequent motion is governed by the coupled differential equations incorporating drag and Magnus effects. Rather than deriving aerodynamic parameters from first principles, we adopt empirical values from the experimental literature to ensure physically realistic serve configurations.
	
	For the spin characteristics of professional serves, we utilize the measurements of Sakurai \textit{et al.}~\cite{sakurai2013spin}, who quantified both spin rates and rotation axis orientations for flat, slice, and kick serves. For the trajectory modifications induced by spin, we employ the computational results of Robinson and Robinson~\cite{robinson2018trajectory}, who solved the equations of motion under realistic aerodynamic conditions for various spin configurations.
	
	The initial ball speeds and angular velocity vectors for the three serve types, expressed in the coordinate system established in Section~2, are:
	\begin{align}
		\text{Flat:} \quad & v_0 = 52.0~\text{m/s}, \quad \boldsymbol{\omega} = (-18.4,\, +22.0,\, +117.0)~\text{rad/s} \\
		\text{Slice:} \quad & v_0 = 46.4~\text{m/s}, \quad \boldsymbol{\omega} = (-73.8,\, +29.6,\, +214.2)~\text{rad/s} \\
		\text{Kick:} \quad & v_0 = 40.8~\text{m/s}, \quad \boldsymbol{\omega} = (-194.1,\, +31.1,\, +267.1)~\text{rad/s}
	\end{align}
	
	The nominal target positions on the service court are defined as $\mathbf{p}_{\text{wide}} = (-4.0,\, 18.29,\, 0)$~m, $\mathbf{p}_{\text{T-zone}} = (-0.1,\, 18.29,\, 0)$~m, and $\mathbf{p}_{\text{body}} = (-2.0,\, 18.29,\, 0)$~m. To compensate for spin-induced trajectory deflection, the aiming points are adjusted: the slice serve target is shifted $+1$~m in the $x$-direction to account for lateral curve, while the kick serve target is extended $+5$~m in the $y$-direction to accommodate the higher, deeper bounce trajectory\cite{robinson2018trajectory}. The resulting aim points are:
	\begin{align}
		\mathbf{p}_{\text{S}} &= (-3.0,\, 18.29,\, 0)~\text{m} \\
		\mathbf{p}_{\text{K}} &= (-0.1,\, 23.29,\, 0)~\text{m} \\
		\mathbf{p}_{\text{F}} &= (-2.0,\, 18.29,\, 0)~\text{m}
	\end{align}
	
	Using the trajectory model of Ref.~\cite{robinson2018trajectory,youn2025} with the empirical spin values, the initial ball velocity vectors required to reach these targets from the impact position are:
	\begin{align}
		\mathbf{v}_{\text{b,F}} &= (-13.66,\, +49.97,\, -4.49)~\text{m/s} \\
		\mathbf{v}_{\text{b,S}} &= (-14.42,\, +43.96,\, -3.49)~\text{m/s} \\
		\mathbf{v}_{\text{b,K}} &= (-5.38,\, +40.43,\, -1.17)~\text{m/s}
	\end{align}
	
	The less negative $v_z$ component in the kick serve reflects the upward initial trajectory necessary for generating the characteristic high-bouncing topspin motion, whereas the flat and slice serves exhibit negative $v_z$ components corresponding to more direct, downward-directed paths.
	
	This empirical approach warrants methodological justification. The aerodynamic behavior of a spinning tennis ball---involving Reynolds number dependence of the drag coefficient and the spin-parameter dependence of the lift coefficient---has been extensively characterized experimentally but remains difficult to predict \textit{ab initio}. By utilizing measured velocity and spin data from actual professional serves~\cite{sakurai2013spin} together with validated trajectory computations incorporating these aerodynamic effects~\cite{robinson2018trajectory}, we obtain initial conditions that reflect realistic serving conditions. This empirical foundation ensures that the subsequent biomechanical analysis addresses physically attainable serve configurations rather than idealized approximations.
	
	\section{Joint Kinematics and Trajectory Optimization}
	
	To transition from the initial serving stance to the impact configuration determined in the previous section, we formulate a boundary value problem for the generalized coordinates $\mathbf{q}(t) \in \mathbb{R}^{n}$, where $n=12$ (three torso angles, three shoulder/clavicle angles, two humeral angles, two wrist angles, and two racket-face angles). For a given serve type (flat, slice, or kick), the desired impact configuration $\mathbf{q}^{(F)}$ and the target racket-head velocity vector $\mathbf{v}_{\text{target}}$ are prescribed. In the implementations shown here, the last two generalized coordinates (racket-face angles) are held fixed, i.e.\ $\theta_{11}(t)\equiv\theta_{11}^{(I)}=\theta_{11}^{(F)}$ and $\theta_{12}(t)\equiv\theta_{12}^{(I)}=\theta_{12}^{(F)}$, so that their polynomial coefficients vanish and they do not participate in the optimization.
	
	\subsection{Temporal Parameterization}
	
	We parameterize the trajectory of the $i$-th generalized coordinate $\theta_i(t)$ using a cubic polynomial time-law. To facilitate the exploration of the swing duration $T$, we introduce a normalized time variable $s = t/T$, where $s \in [0, 1]$. The angular trajectory is defined as
	\begin{equation}
		\theta_i(s) = \theta_i^{(I)} + c_{1,i}\, s + c_{2,i}\, s^2 + c_{3,i}\, s^3 ,
		\label{eq:cubic_norm}
	\end{equation}
	where $\theta_i^{(I)}$ is the initial angle, and $\{c_{1,i}, c_{2,i}, c_{3,i}\}$ are the polynomial coefficients to be determined. The time derivatives (angular velocity $\dot{\theta}_i$ and acceleration $\ddot{\theta}_i$) are related to the derivatives with respect to $s$ by the scaling factor $T$:
	\begin{equation}
		\dot{\theta}_i(t) = \frac{1}{T} \frac{d\theta_i}{ds}, \qquad 
		\ddot{\theta}_i(t) = \frac{1}{T^2} \frac{d^2\theta_i}{ds^2} .
	\end{equation}
	
	\subsection{Constraint Formulation}
	
	The determination of the coefficients constitutes a constrained optimization problem. The system is subject to the following boundary conditions:
	
	1. \textbf{Initial configuration:} By construction of Eq.~(\ref{eq:cubic_norm}), $\theta_i(0) = \theta_i^{(I)}$ is automatically satisfied.
	
	2. \textbf{Final configuration:} At impact ($s=1$), the joint angles must reach the target configuration $\theta_i^{(F)}$ derived from the inverse-kinematics analysis:
	\begin{equation}
		c_{1,i} + c_{2,i} + c_{3,i} = \Delta \theta_i,
		\label{eq:pos_constraint}
	\end{equation}
	where $\Delta \theta_i = \theta_i^{(F)} - \theta_i^{(I)}$. For the frozen racket-face joints ($i=11,12$) we have $\Delta\theta_i = 0$ and set $c_{1,i}=c_{2,i}=c_{3,i}=0$.
	
	3. \textbf{Impact velocity constraint:} The kinematic chain must generate a specific linear velocity vector $\mathbf{v}_{\text{target}}$ at the racket-head center at the moment of impact. This imposes a constraint on the generalized velocities $\dot{\mathbf{q}}(T)$:
	\begin{equation}
		\mathbf{J}(\mathbf{q}^{(F)}) \cdot \dot{\mathbf{q}}(T) = \mathbf{v}_{\text{target}},
		\label{eq:vel_constraint_vec}
	\end{equation}
	where $\mathbf{J}(\mathbf{q}) \in \mathbb{R}^{3 \times n}$ is the translational Jacobian matrix of the racket-head center, evaluated numerically by finite differences of the forward-kinematics model at $\mathbf{q}^{(F)}$. Using Eq.~(\ref{eq:cubic_norm}), the components of the generalized velocity at impact are
	\[
	\dot{\theta}_i(T) = \frac{1}{T}\Bigl(c_{1,i} + 2c_{2,i} + 3c_{3,i}\Bigr),
	\]
	so that Eq.~(\ref{eq:vel_constraint_vec}) becomes a linear condition on the coefficient vectors.
	
	\subsection{KKT Optimization and Solution}
	
	The polynomial coefficients are not uniquely determined by these constraints: for $n$ generalized coordinates there are in principle $3n$ coefficients, whereas the final position constraint Eq.~(\ref{eq:pos_constraint}) and the three components of Eq.~(\ref{eq:vel_constraint_vec}) provide only $n+3$ scalar conditions (and two coordinates are frozen). To resolve this redundancy, we introduce a quadratic regularization that penalizes large coefficients and thereby favors dynamically economical trajectories.
	
	Eliminating $c_{3,i}$ using the position constraint Eq.~(\ref{eq:pos_constraint}),
	\[
	c_{3,i} = \Delta\theta_i - c_{1,i} - c_{2,i},
	\]
	the angular velocity at impact can be written as
	\[
	\dot{\theta}_i(T) = \frac{1}{T}\bigl(3\Delta\theta_i - 2c_{1,i} - c_{2,i}\bigr),
	\]
	or in vector form
	\[
	\dot{\boldsymbol{\theta}}(T) = \frac{1}{T}\bigl(3\Delta\boldsymbol{\theta} - 2\mathbf{c}_1 - \mathbf{c}_2\bigr),
	\]
	where $\mathbf{c}_1 = (c_{1,1},\ldots,c_{1,n})^\top$ and $\mathbf{c}_2 = (c_{2,1},\ldots,c_{2,n})^\top$. For a fixed swing duration $T$, we then solve
	\[
	\min_{\mathbf{c}_1,\mathbf{c}_2} \ \|\mathbf{c}_1\|_2^2 + \|\mathbf{c}_2\|_2^2
	\quad\text{subject to}\quad
	\mathbf{J}(\mathbf{q}^{(F)})\cdot\dot{\boldsymbol{\theta}}(T) = \mathbf{v}_{\text{target}} .
	\]
	This constrained quadratic problem is solved using Lagrange multipliers, yielding a linear Karush–Kuhn–Tucker (KKT) system for $\mathbf{c}_1$, $\mathbf{c}_2$ and the three multipliers enforcing the components of the velocity constraint.
	
	In practice, $T$ is treated as a scalar design parameter constrained to a physiologically plausible interval. For any $T$ in this range, the KKT system admits a unique solution that satisfies Eq.~(\ref{eq:vel_constraint_vec}) to numerical precision, so that the residual
	\[
	\left\| \mathbf{J}(\mathbf{q}^{(F)}) \cdot \dot{\mathbf{q}}(T; \mathbf{c}^*) - \mathbf{v}_{\text{target}} \right\|
	\]
	is essentially flat as a function of $T$. Consequently, the kinematic constraints alone do not identify a unique optimal duration; we select $T$ within the interval typically reported for professional serves (about $0.25$–$0.35$~s). For the modeled serves, the chosen durations lie in this range and reproduce the target racket-head velocity to within numerical tolerance.
	Table~\ref{tab:joint_angles} lists final angles for each serve type.
	
	\begin{table}[htbp]
		\centering
		\caption{Final joint angles (degrees) for different serve types.}
		\label{tab:joint_angles}
		\begin{tabular}{lccc}
			\toprule
			Joint & Flat (Final) & Kick (Final) & Slice (Final) \\
			\midrule
			$\theta_1$  & -25.00 & -25.05 & -25.35 \\
			$\theta_2$  &   5.00 &  10.79 &   8.42 \\
			$\theta_3$  &  24.89 &  10.00 &  30.00 \\
			$\theta_4$  &  22.57 &  19.49 &  22.84 \\
			$\theta_5$  &  -4.91 &  10.95 &  13.18 \\
			$\theta_6$  &  78.00 &  55.76 &  77.96 \\
			$\theta_7$  & 175.16 & 173.33 & 180.00 \\
			$\theta_8$  & -22.69 &  -7.43 & -27.97 \\
			$\theta_9$  & -40.04 & -59.93 & -49.80 \\
			$\theta_{10}$ & 34.88 & 26.63 & 10.01 \\
			$\theta_{11}$ & 50.00 & 50.00 & 50.00 \\
			$\theta_{12}$ & 30.00 & 30.00 & 30.00 \\
			\bottomrule
		\end{tabular}
	\end{table}
	
	These trajectories achieve racket normals with dot products to the prescribed target directions exceeding $0.99$, confirming kinematic fidelity of the impact configuration.

	\begin{figure}[htbp]
		\centering
		\includegraphics[width=10cm]{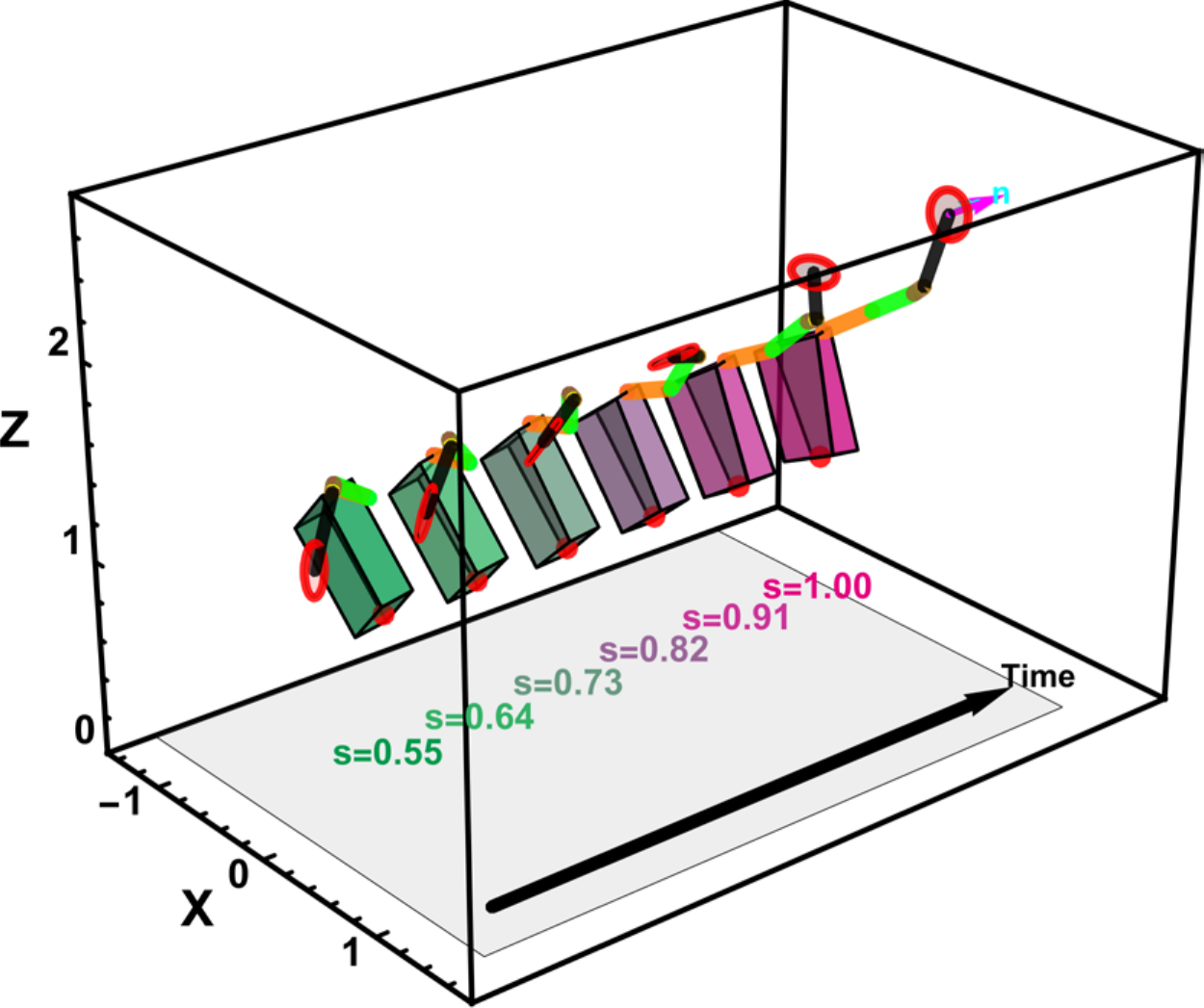}
		\caption{Sequential illustration of the tennis flat serve motion from the preparation phase to the moment of impact. The vectors $\vec{n}$  and  $\vec{v}$  denote the normal vector and velocity vector of the racket head, respectively. Here, $s$ represents the  normalized  time elapsed from the preparation phase. } \label{flatServe3D}
	\end{figure}
	\begin{figure}[htbp]
		\centering
		\includegraphics[width=10cm]{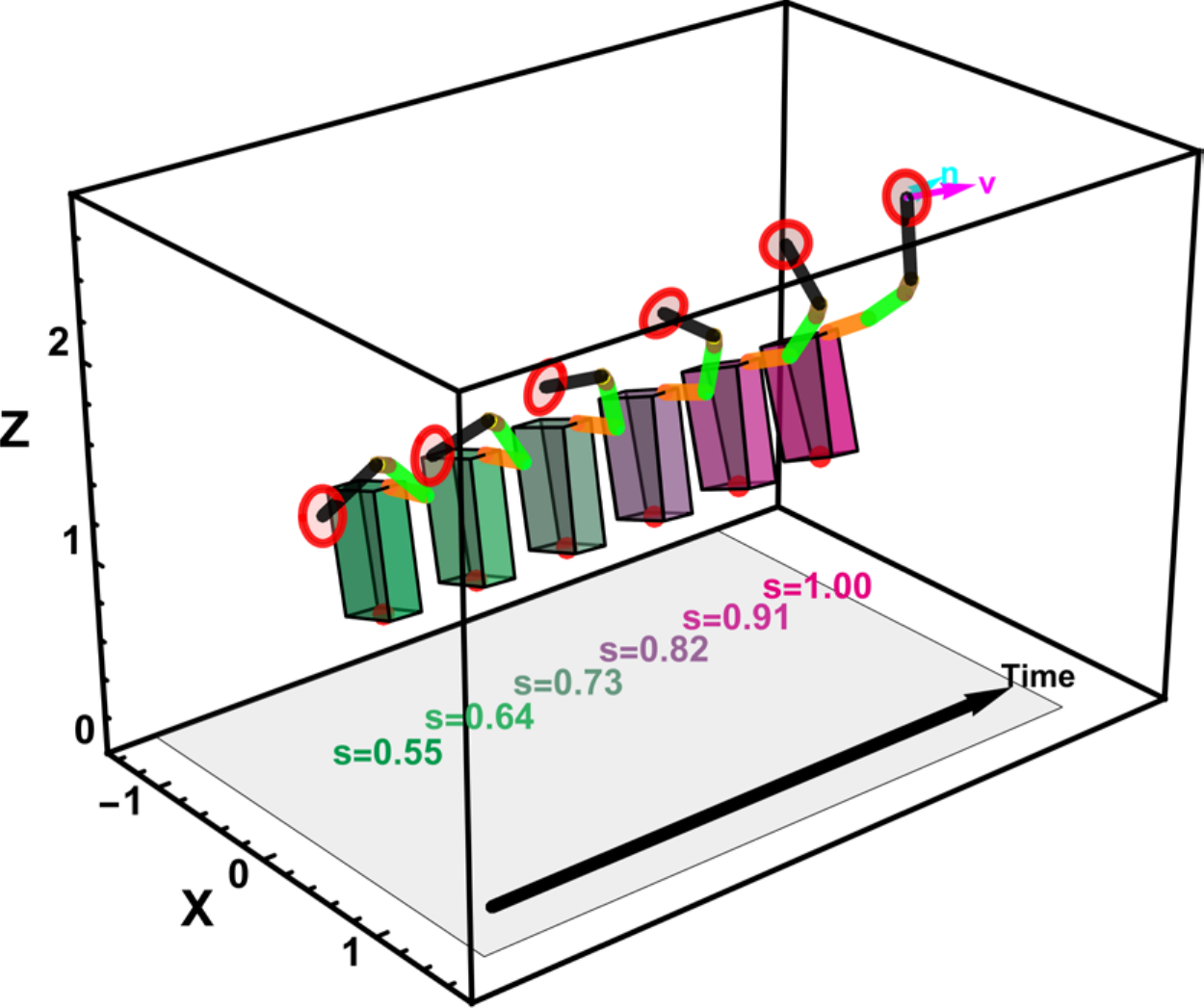}
		\caption{Sequential illustration of the tennis slice serve motion from the preparation phase to the moment of impact. The vectors $\vec{n}$  and  $\vec{v}$  denote the normal vector and velocity vector of the racket head, respectively. Here, $s$ represents the normalized  time elapsed from the preparation phase.  } \label{sliceServe3D}
	\end{figure}
	\begin{figure}[htbp]
		\centering
		\includegraphics[width=10cm]{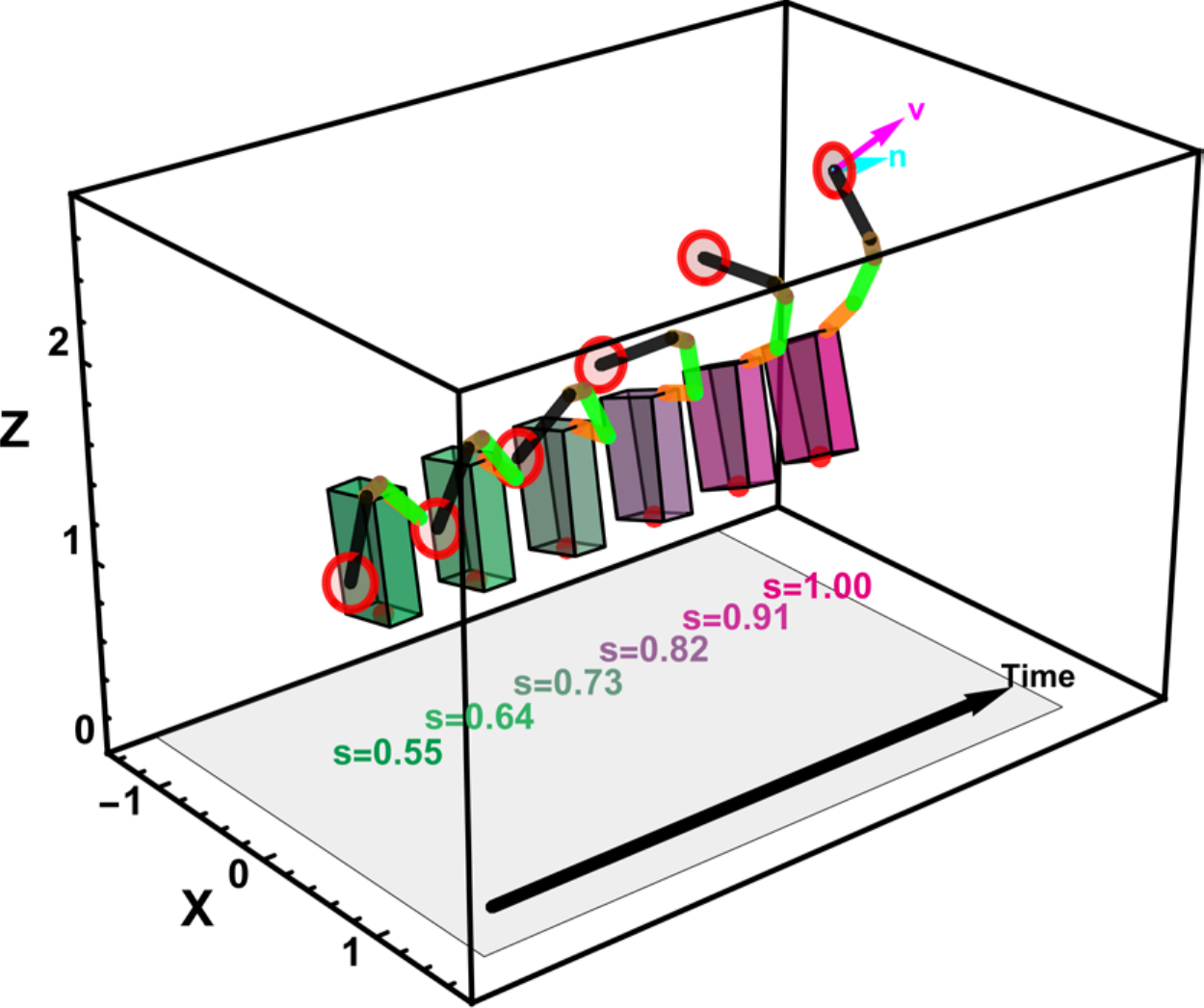}
		\caption{Sequential illustration of the tennis kick serve motion from the preparation phase to the moment of impact. The vectors $\vec{n}$  and  $\vec{v}$  denote the normal vector and velocity vector of the racket head, respectively. Here, $s$ represents the  normalized  time elapsed from the preparation phase.  } \label{kickServe3D}
	\end{figure}
	\begin{figure}[htbp]
		\centering
		\includegraphics[width=10cm]{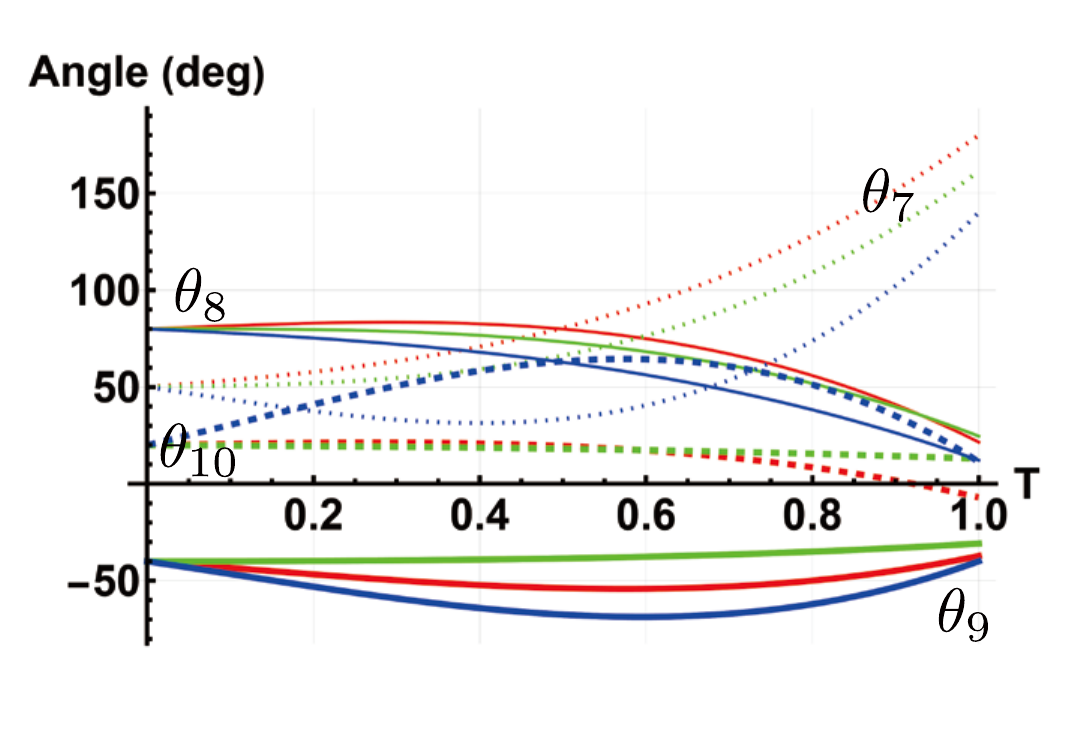}
		\caption{Time dependence of arm and wrist joint angles ($\theta_7$--$\theta_{10}$) for flat (red), slice (green), and kick (blue) serves. Line styles indicate specific angles: $\theta_7$ (thin solid), $\theta_8$ (thin dotted), $\theta_9$ (thick solid), and $\theta_{10}$ (thick dashed). The time axis is normalized to the impact moment ($T=1$). } \label{Angle7To10}
	\end{figure}

	In this section, we employed optimization techniques to determine specific temporal evolutions of the joint angles for three types of serves. These solutions satisfy the conditions derived in the previous section regarding the angle between the racket's velocity vector and the normal vector of the racket face at the moment of impact. Table~\ref{tab:joint_angles} presents the values of the joint angles at impact, while Figures \ref{flatServe3D}, \ref{sliceServe3D}, and \ref{kickServe3D} illustrate the three-dimensional temporal evolution of the serve motions.
	
	For the flat serve depicted in Figure \ref{flatServe3D}, it is observed that the normal direction of the racket head aligns perfectly with the direction of the racket's velocity vector at the instant of impact ($s=1$). In comparison, the slice serve motion shown in Figure \ref{sliceServe3D} exhibits a trajectory where the racket does not ascend directly parallel to the back; instead, it moves further away from the torso axis before swinging around toward the target point. Since the slice serve requires a deviation between the normal vector of the racket face and the velocity vector at impact, the orientation of the racket face is distinct from its propagation direction as it approaches the impact point. Furthermore, the kick serve motion illustrated in Figure \ref{kickServe3D} reveals that, compared to the slice serve, the racket remains below the shoulder level for a longer duration before ascending rapidly as it approaches the impact point. Consequently, the direction of the racket's velocity vector at impact is directed significantly more upwards than in the slice or flat serves.
	
	Although the results at the impact point are an expected consequence of the boundary conditions imposed on the racket velocity and normal vectors during the optimization simulation, analyzing the approach trajectory from the loading phase to achieve these conditions provides critical insights for tennis serve mechanics.
	
	Figure \ref{Angle7To10} displays the temporal evolution of the joint angles ($\theta_7 \dots \theta_{10}$), which exhibit significant variation, to highlight the distinct characteristics of the three serve types. This figure illustrates the trajectories of these joint angles from an identical initial position to the point of impact. The flat, slice, and kick serves are represented in red, green, and blue, respectively, with the time normalized such that the moment of impact corresponds to $T=1$. These plots effectively demonstrate the time-dependent range of motion for the joint angles across the different serve types. However, it should be noted that a constant joint angle does not necessarily imply that the torque acting on that joint is constant. To investigate this relationship, we will examine the joint torques and forces in the following section.

	\section{Inverse Dynamics Analysis of Joint Torques}
	
	To elucidate the mechanical loading required to execute the optimized kinematic trajectories, we perform an inverse-dynamics analysis. We adopt a hybrid formalism that combines the classical Newton–Euler equations for rigid-body dynamics with the Principle of Virtual Power to project spatial wrenches onto the generalized coordinates.
	
	\subsection{Rigid Body Dynamics in $\mathbb{R}^3$}
	
	The multi-segment human model comprises $N_b=5$ rigid bodies (torso, clavicle, upper arm, hand, and racket). For each body $k$, the instantaneous kinematics (linear acceleration $\mathbf{a}_{G,k}$ of the center of mass and angular acceleration $\boldsymbol{\alpha}_k$) are computed via numerical differentiation of the forward kinematics, utilizing a central-difference scheme with a time step $h=2$~ms~\cite{winter2009biomechanics}. This choice provides a good compromise between temporal resolution and robustness to high-frequency numerical noise.
	
	The effective net force $\mathbf{F}_k$ and net moment $\mathbf{N}_k$ associated with the center of mass of the $k$-th body are governed by the Newton–Euler equations~\cite{featherstone2008rigid}
	\begin{align}
		\mathbf{F}_k &= m_k \bigl(\mathbf{a}_{G,k} - \mathbf{g}\bigr), \\
		\mathbf{N}_k &= \mathbf{I}_k \cdot \boldsymbol{\alpha}_k 
		+ \boldsymbol{\omega}_k \times \bigl(\mathbf{I}_k \cdot \boldsymbol{\omega}_k\bigr),
	\end{align}
	where $m_k$ is the segment mass, $\mathbf{g}$ is the gravitational acceleration vector, $\mathbf{I}_k$ is the inertia tensor expressed in the world frame, and $\boldsymbol{\omega}_k$ is the angular velocity vector. The inertia tensor is updated at each time step via the similarity transformation
	\begin{equation}
		\mathbf{I}_k(t) = \mathbf{R}_k(t)\, \mathbf{I}_k^{\text{body}}\, \mathbf{R}_k(t)^\top ,
	\end{equation}
	where $\mathbf{R}_k(t)$ is the rotation matrix from the body-fixed frame to the world frame and $\mathbf{I}_k^{\text{body}}$ is the constant inertia in body coordinates~\cite{zatsiorsky2002kinetics}.
	
	\subsection{Generalized Torques via Virtual Power}
	
	In a system of coupled rigid bodies, the contact and joint forces are not directly known. To extract the generalized joint torques $\boldsymbol{\tau} \in \mathbb{R}^n$ corresponding to the active muscular moments, we invoke the Principle of Virtual Power. The total mechanical power of the system must equal the sum of the scalar products of the effective inertial wrenches and the twist velocities of each body.
	
	The generalized torque $\tau_i$ associated with the $i$-th joint is obtained by projecting the net spatial forces and moments of all bodies onto the motion subspace of that joint. This is mathematically expressed using the translational Jacobian $\mathbf{J}_{v,k}$ and the rotational Jacobian $\mathbf{J}_{\omega,k}$ for each body $k$:
	\begin{equation}
		\boldsymbol{\tau} = \sum_{k=1}^{N_b} \left[ \mathbf{J}_{v, k}^\top \cdot \mathbf{F}_k 
		+ \mathbf{J}_{\omega, k}^\top \cdot \mathbf{N}_k \right],
		\label{eq:projection}
	\end{equation}
	where the Jacobians relate the generalized velocities to the body twists via
	\[
	\mathbf{v}_{G,k} = \mathbf{J}_{v, k} \, \dot{\mathbf{q}}, 
	\qquad
	\boldsymbol{\omega}_k = \mathbf{J}_{\omega, k} \, \dot{\mathbf{q}} .
	\]
	In the implementation, both $\mathbf{J}_{v,k}$ and $\mathbf{J}_{\omega,k}$ are obtained by finite-difference differentiation of the forward-kinematics functions with respect to the generalized coordinates.
	
	\subsection{Torque Decomposition: Inertial vs. Gravitational}
	
	The linearity of the projection operator in Eq.~(\ref{eq:projection}) allows for the decomposition of the total torque into an inertial component $\boldsymbol{\tau}_{\text{in}}$, required to accelerate the masses, and a gravitational component $\boldsymbol{\tau}_{\text{g}}$, required to support the bodies against gravity:
	\begin{align}
		\boldsymbol{\tau}_{\text{in}} &= \sum_{k=1}^{N_b} \left[ \mathbf{J}_{v, k}^\top \cdot (m_k \mathbf{a}_{G,k}) 
		+ \mathbf{J}_{\omega, k}^\top \cdot \bigl(\mathbf{I}_k \boldsymbol{\alpha}_k 
		+ \boldsymbol{\omega}_k \times \mathbf{I}_k \boldsymbol{\omega}_k\bigr) \right], \\
		\boldsymbol{\tau}_{\text{g}} &= -\sum_{k=1}^{N_b} \mathbf{J}_{v, k}^\top \cdot (m_k \mathbf{g}) .
	\end{align}
	This decomposition provides physical insight into the serving mechanism. Our analysis reveals that during the acceleration phase ($t/T > 0.7$), the inertial terms dominate, particularly for the humeral pitch and axial rotation degrees of freedom and for the wrist joints, indicating that the modeled serves are predominantly ballistic motions driven by dynamic transfer of momentum rather than by static gravitational loading.

	\begin{figure}[htbp]
		\centering
		\includegraphics[width=10cm]{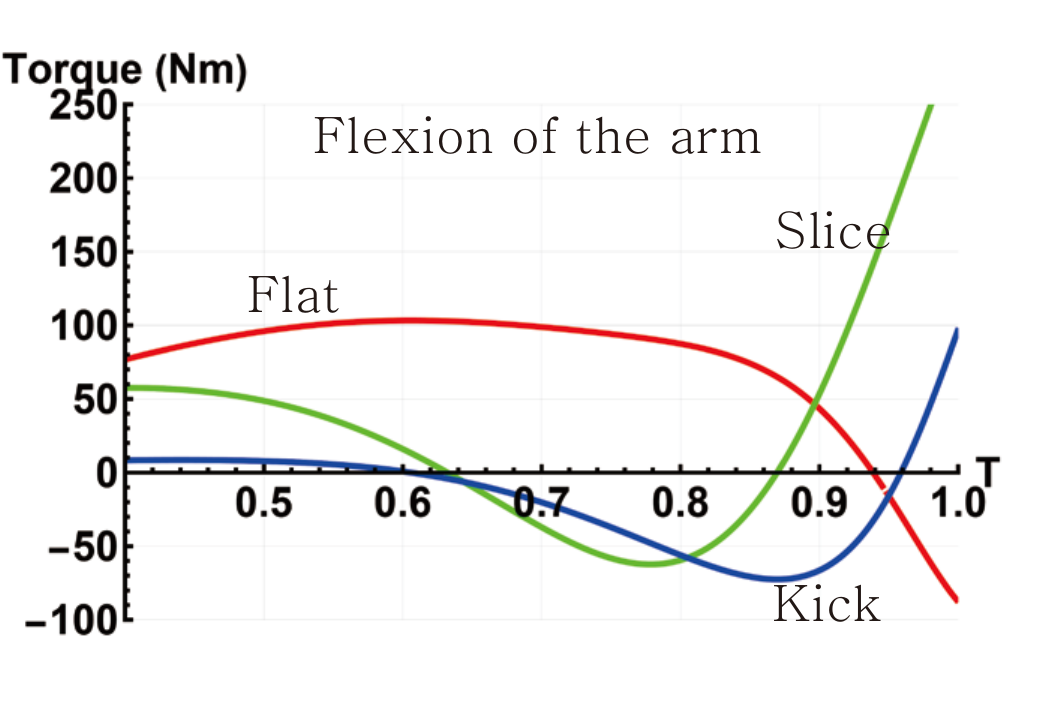}
		\caption{Temporal evolution of the applied torques for the three types of tennis serves. The figures show the torques required to generate the effective flexion/extension of the distal arm segment relative to the shoulder chain ($\theta_7$).
		} \label{torque7}
	\end{figure}
	\begin{figure}[htbp]
		\centering
		\includegraphics[width=10cm]{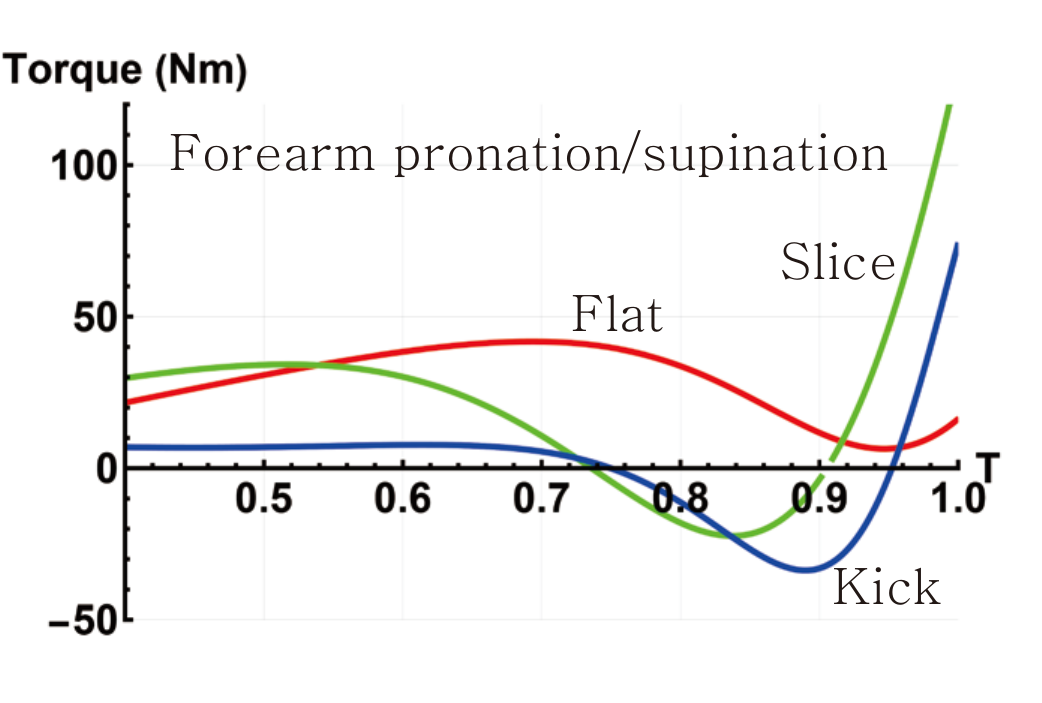}
		\caption{Temporal evolution of the applied torques for the three types of tennis serves. The figures show the torques required to generate
			the axial twist of the distal arm segment about its longitudinal axis ($\theta_8$).
		} \label{torque8}
	\end{figure}
	\begin{figure}[htbp]
		\centering
		\includegraphics[width=10cm]{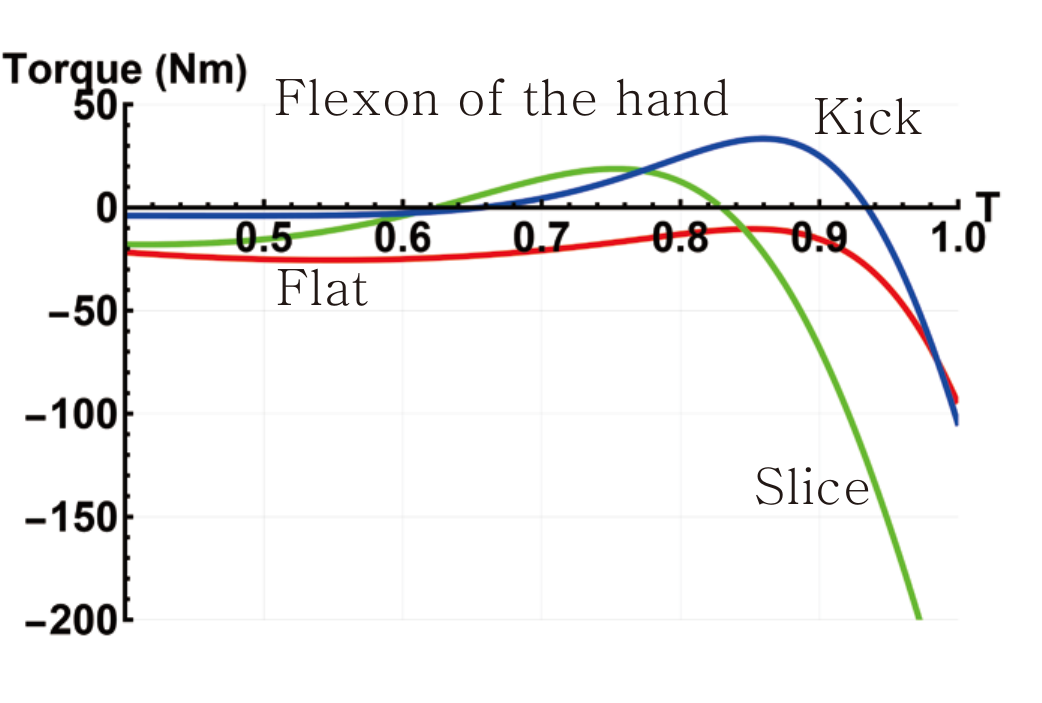}
		\caption{Temporal evolution of the applied torques for the three types of tennis serves. The figures show the torques required to generate
			the flexion/extension of the hand relative to the distal arm segment ($\theta_9$). } \label{torque9}
	\end{figure}

	\begin{figure}[htbp]
		\centering
		\includegraphics[width=10cm]{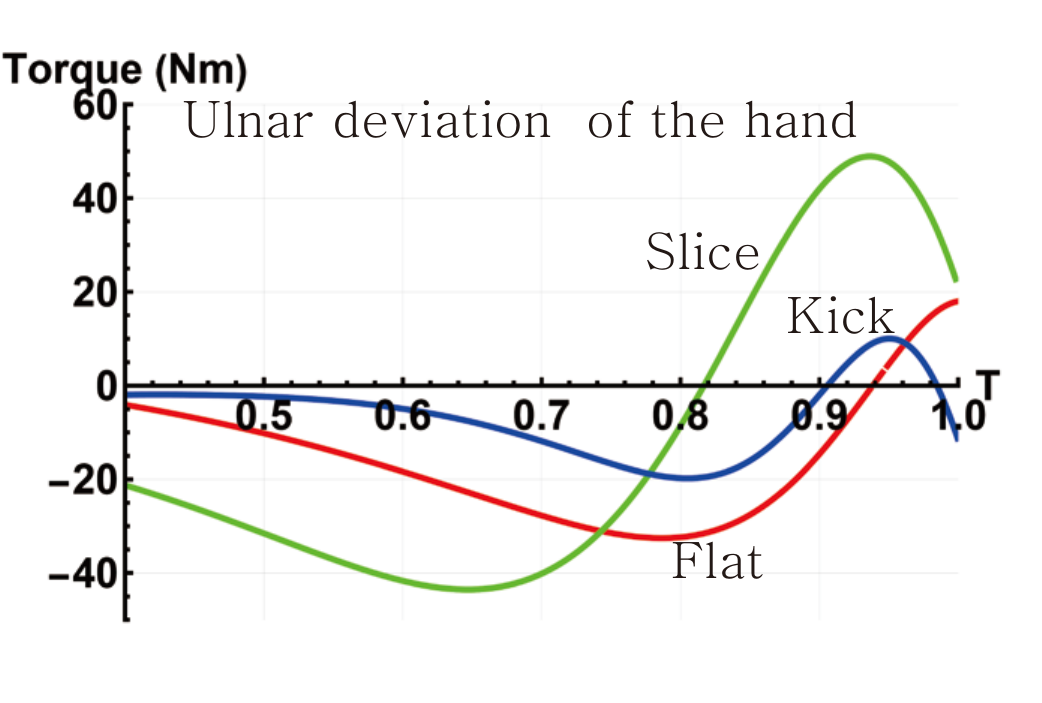}
		\caption{Temporal evolution of the applied torques for the three types of tennis serves. The figures show the torques required to generate the radial/ulnar deviation of the hand relative to the distal arm segment ($\theta_{10}$). } \label{torque10}
	\end{figure}
	
	In this section, the joint torques required during the three types of serves are computed from the time histories of the generalized coordinates obtained in the previous section. In principle, the generalized forces (and thus the joint torques) can be derived from the Lagrange equations. However, for a system involving many joints, the number of variables grows rapidly, making an analytic derivation of the generalized forces intractable. Therefore, the torques were instead calculated using the Newton-Euler equations for rigid-body dynamics in combination with the Principle of Virtual Power.
	
	As seen in Figure~\ref{Angle7To10}, the temporal trends of the joint angles are broadly similar across the three serves, with differences mainly in the magnitude of the variations. In contrast, the torque time histories obtained in this section show pronounced differences. The torque required to change a particular joint angle is not determined solely by the direct muscular effort at that joint; it must also counteract gravity acting on all body segments and the racket, as well as inertial and interaction forces arising from the three-dimensional motion of all joints.
	
	Figure~\ref{torque7} shows the torque about $\theta_7$ (effective flexion/extension of the distal arm segment relative to the shoulder chain), including the effects of gravity and full spatial motion, that is required to reproduce the prescribed angular trajectory during the serve. For the flat serve, an almost constant torque is applied up to just before impact, whereupon the torque magnitude drops abruptly and reverses sign near the impact point. For the slice and kick serves, by contrast, the torque about $\theta_7$ remains small for most of the motion and then increases sharply near impact. If one were to observe only the time evolution of $\theta_7$ from outside, all three serves would appear to exhibit a similarly increasing trend. However, once the three-dimensional motion and the coupling with the other joints are taken into account, it becomes evident that the torques required about $\theta_7$ differ substantially among the three serves.
	
	Figures~\ref{torque9} and \ref{torque10} depict the torque variations associated with wrist flexion/extension ($\theta_9$) and radial/ulnar deviation ($\theta_{10}$) of the hand relative to the distal arm segment. For $\theta_9$, both the angular trajectories and the torque profiles are very similar across the three serve types. In contrast, for $\theta_{10}$, the slice and kick serves exhibit almost the same angular trend over time, while the kick serve shows a noticeably different path near impact. Nevertheless, the torque about $\theta_{10}$ is actually largest and most distinct in the slice serve compared with the other serves. This indicates that the time evolution of a given joint angle is influenced not only by the torque applied locally at that joint but also by gravity and by the spatial positions and motions of all other joints. Consequently, attempting to achieve a specific angular change at one joint by acting only on that joint may in fact interfere with the coordination of the overall movement.
	
	From Figure~\ref{Angle7To10}, we see that $\theta_9$ and $\theta_{10}$ remain almost unchanged from the preparation phase up to the instant of ball impact. However, Figs.~\ref{torque9} and \ref{torque10} show that, for the slice serve (green curves), the torques required about $\theta_9$ and $\theta_{10}$ exhibit the largest and most pronounced variations among the three serves. This leads to an important conclusion for serve analysis based on photographs or external observation: the absence of visible changes in a joint angle does not imply that no torque is being applied at that joint. Care must therefore be taken not to infer the underlying joint kinetics solely from observed kinematics.

	\section{Summary and Discussion}
	
	This study presented a complete physics-based modeling pipeline that connects empirically observed professional tennis serve characteristics to the underlying joint-level mechanics. Starting from measured ball spin and velocity data of elite players, we first determined realistic initial velocity vectors and racket-face orientations required for flat, slice, and kick serves to land in designated target zones under real aerodynamic conditions. These terminal constraints were then used as boundary conditions for a 12-DOF upper-body kinematic chain.
	
	By parameterizing joint trajectories with minimal-degree polynomials and solving a constrained quadratic optimization problem, we obtained smooth, physiologically plausible joint angle time histories that exactly satisfy the prescribed impact configuration and racket-head velocity while minimizing unnecessary angular excursions. Unlike traditional video-based motion capture studies that report only what players \textit{do}, the present approach elucidates \textit{how} a desired racket state can be achieved analytically through coordinated multi-joint motion. The resulting three-dimensional animations (Figs.~\ref{flatServe3D}--\ref{kickServe3D}) and joint-angle trajectories (e.g., Figure~\ref{Angle7To10}) reveal clear kinematic distinctions: the kick serve delays upward acceleration of the racket until late in the swing, whereas the slice serve exhibits greater lateral excursion of the racket path.
	
	The subsequent inverse dynamics analysis demonstrated that even joints exhibiting nearly constant angles during large portions of the motion (particularly the wrist degrees of freedom $\theta_9$ and $\theta_{10}$) require substantial and highly time-varying torques (Figs.~\ref{torque9} and \ref{torque10}). This counterintuitive result arises from gravitational loading, Coriolis effects, and dynamic coupling across the entire kinematic chain. A practical implication is that visual feedback of joint angles alone is insufficient for motor learning or correction~\cite{reid2016biomechanics}: players must develop internal representations of the required torque profiles to maintain stability when distal segments appear “quiet.”
	
	It must be emphasized that the trajectories derived here represent one family of solutions that minimize a simple quadratic regularity condition on the polynomial coefficients. They are not claimed to be globally optimal with respect to metabolic energy, peak joint power, or robustness to perturbation. Numerous other coordination patterns undoubtedly exist that achieve the same terminal racket state with different intermediate kinematics. The present work therefore does not prescribe “the correct technique,” but rather illustrates a rigorous, reproducible method by which any desired serve outcome can be reverse-engineered into joint-space commands.
	
	In contrast to contemporary machine-learning approaches that directly map desired ball trajectories to final body postures or muscle activation patterns, the current framework retains full transparency at every stage: each torque contribution can be traced back to specific inertial, gravitational, or centrifugal terms. This interpretability is particularly valuable for coaching, injury prevention, and rehabilitation~\cite{elliott2003technique, kibler2013shoulder}, where understanding \textit{why} certain joints experience high loading is often more important than merely replicating an elite player’s external motion.
	
	Future extensions of this model could incorporate closed-loop neuromuscular control~\cite{delp2007opensim}, elastic tendon dynamics, or active adjustment of the hip position and leg drive, which are known to contribute significantly to total racket speed in professional serves. Nevertheless, even in its current upper-body-focused form, the model successfully bridges the gap between aerodynamic outcome and musculoskeletal cause, providing a physically grounded reference for both scientific analysis and practical training of the tennis serve.



\begin{thebibliography}{99}
		
		\bibitem{elliott2006biomechanics}
		Elliott, B. (2006).
		Biomechanics and tennis.
		\textit{British Journal of Sports Medicine}, 40(5), 392--396.
		\href{https://doi.org/10.1136/bjsm.2005.023150}{doi:10.1136/bjsm.2005.023150}
		
		\bibitem{fleisig2003kinematics}
		Fleisig, G., Nicholls, R., Elliott, B., \& Escamilla, R. (2003).
		Kinematics used by world class tennis players to produce high-velocity serves.
		\textit{Sports Biomechanics}, 2(1), 51--64.
		\href{https://doi.org/10.1080/14763140308522807}{doi:10.1080/14763140308522807}
		
		\bibitem{cust2019machine}
		Cust, E. E., Sweeting, A. J., Ball, K., \& Robertson, S. (2019).
		Machine and deep learning for sport-specific movement recognition: A systematic review of model development and performance.
		\textit{Journal of Sports Sciences}, 37(5), 568--600.
		\href{https://doi.org/10.1080/02640414.2018.1521769}{doi:10.1080/02640414.2018.1521769}
		
		\bibitem{youn2015}
		Youn, S.-H. (2015).
		Double pendulum model for a tennis stroke including a collision process.
		\textit{Journal of the Korean Physical Society}, 67(7), 1110--1117.
		\href{https://doi.org/10.3938/jkps.67.1110}{doi:10.3938/jkps.67.1110}
		
		\bibitem{youn2023}
		Youn, S.-H. (2023).
		Push vs. hit, the most efficient stroke in the kinetic chain process.
		\textit{Journal of the Korean Physical Society}, 82(7), 629--637.
		\href{https://doi.org/10.1007/s40042-023-00751-8}{doi:10.1007/s40042-023-00751-8}
		
		\bibitem{youn2017}
		Youn, S.-H. (2017).
		Study on the Motion of a Tennis Ball for an Efficient Swing Pattern to Send the Ball Toward a Target.
		\textit{New Physics: Sae Mulli}, 67(11), 1378--1387.
		\href{https://doi.org/10.3938/NPSM.67.1378}{doi:10.3938/NPSM.67.1378}
		
		\bibitem{youn2025}
		Youn, S. H. (2025).
		Spin generation in tennis ball-racket collisions.
		\textit{New Physics: Sae Mulli}, 75(7), 543--555.
		
		\bibitem{brody2002physics}
		Brody, Howard. The physics of tennis. III. The ball–racket interaction. 	\textit{American Journal of Physics} 65.10 (1997): 981-987.\href{https://doi.org/10.1119/1.18701}{doi:10.1119/1.18701}
		
		\bibitem{cross2011physics}
		Cross, R. (2011).
		\textit{Physics of Baseball and Softball}.
		Springer New York.
		\href{https://doi.org/10.1007/978-1-4419-8113-4}{doi:10.1007/978-1-4419-8113-4}
		
		\bibitem{cross2003oblique}
		Cross, R. (2003).
		Oblique impact of a tennis ball on the strings of a tennis racket.
		\textit{Sports Engineering}, 6(4), 235--254.
		\href{https://doi.org/10.1007/BF02844026}{doi:10.1007/BF02844026}
		
		\bibitem{goodwill2004ball}
		Goodwill, S. R., \& Haake, S. J. (2004).
		Ball spin generation for oblique impacts with a tennis racket.
		\textit{Experimental Mechanics}, 44(2), 195--206.
		\href{https://doi.org/10.1007/BF02428179}{doi:10.1007/BF02428179}
		
		\bibitem{sprigings1994scientific}
		Sprigings, E., Marshall, R., Elliott, B., \& Jennings, L. (1994).
		A three-dimensional kinematic method for determining the effectiveness of arm segment rotations in producing racquet-head speed.
		\textit{Journal of Biomechanics}, 27(3), 245--254.
		\href{https://doi.org/10.1016/0021-9290(94)90001-9}{doi:10.1016/0021-9290(94)90001-9}
		
		\bibitem{mehta2001sports}
		Mehta, Rabindra D. Sports ball aerodynamics. \textit{Sport aerodynamics}. 
		Vienna: Springer Vienna, 2008. 229-331. \href{https://doi.org/10.1007/978-3-211-89297-8\_12}{doi:10.1007/978-3-211-89297-8\_12}
		
		\bibitem{sakurai2013spin}
		Sakurai, S., Reid, M., \& Elliott, B. (2013).
		Ball spin in the tennis serve: Spin rate and axis of rotation.
		\textit{Sports Biomechanics}, 12(1), 23--29.
		\href{https://doi.org/10.1080/14763141.2012.671355}{doi:10.1080/14763141.2012.671355}
		
		\bibitem{robinson2018trajectory}
		Robinson, G., \& Robinson, I. (2018).
		Model trajectories for a spinning tennis ball: I. The service stroke.
		\textit{Physica Scripta}, 93(12), 123002.
		\href{https://doi.org/10.1088/1402-4896/aae733}{doi:10.1088/1402-4896/aae733}
		
		\bibitem{winter2009biomechanics}
		Winter, D. A. (2009).
		\textit{Biomechanics and Motor Control of Human Movement} (4th ed.).
		John Wiley \& Sons.
		\href{https://doi.org/10.1002/9780470549148}{doi:10.1002/9780470549148}
		
		\bibitem{featherstone2008rigid}
		Featherstone, R. (2008).
		\textit{Rigid Body Dynamics Algorithms}.
		Springer US.
		\href{https://doi.org/10.1007/978-1-4899-7560-7}{doi:10.1007/978-1-4899-7560-7}
		
		\bibitem{zatsiorsky2002kinetics}
		Zatsiorsky, V. M. (2002).
		\textit{Kinetics of Human Motion}.
		Human Kinetics.
		ISBN: 978-0880116763
		
		\bibitem{reid2016biomechanics}
		Reid, M., Elliott, B., \& Crespo, M. (2013).
		Mechanics and learning practices associated with the tennis forehand: A review.
		\textit{Journal of Sports Science and Medicine}, 12(2), 225--231.
		\href{https://pmc.ncbi.nlm.nih.gov/articles/PMC3761830/} {PMID: 24149800}
		
		\bibitem{elliott2003technique}
		Elliott, B., Fleisig, G., Nicholls, R., \& Escamilla, R. (2003).
		Technique effects on upper limb loading in the tennis serve.
		\textit{Journal of Science and Medicine in Sport}, 6(1), 76--87.
		\href{https://doi.org/10.1016/S1440-2440(03)80011-7}{doi:10.1016/S1440-2440(03)80011-7}
		
		\bibitem{kibler2013shoulder}
		Kibler, W. B., Wilkes, T., \& Sciascia, A. (2013).
		Mechanics and pathomechanics in the overhead athlete.
		\textit{Clinics in Sports Medicine}, 32(4), 637--651.
		\href{https://doi.org/10.1016/j.csm.2013.07.003}{doi:10.1016/j.csm.2013.07.003}
		
		\bibitem{delp2007opensim}
		Delp, S. L., Anderson, F. C., Arnold, A. S., Loan, P., Habib, A., John, C. T., ... \& Thelen, D. G. (2007).
		OpenSim: Open-source software to create and analyze dynamic simulations of movement.
		\textit{IEEE Transactions on Biomedical Engineering}, 54(11), 1940--1950.
		\href{https://doi.org/10.1109/TBME.2007.901024}{doi:10.1109/TBME.2007.901024}
		
		
	\end{thebibliography}
\end{document}